\documentclass[aps]{revtex4}
\usepackage{bm}
\usepackage{graphicx,color}
\usepackage{subfigure}
\usepackage{float}
\def\<{{\langle}}
\def\>{{\rangle}}

\def\p{{\mathbf p}}

\def\k{{\mathbf k}}

\def\n{{\mathbf n}}

\def\vsigma{{\bm\sigma}}
\begin{document}
\title{Numerical and Physical Stability of Supernova Neutrino Flavor Evolution}
\author{B. D. Keister}
\affiliation{
Physics Division, National Science Foundation, Arlington, VA 22230}
\begin{abstract}
This paper examines neutrino flavor evolution outside a supernova neutrinosphere using a one-dimensional model that retains the non-linear nature of neutrino-neutrino interactions as well as some aspects of the full geometry.  In some limiting cases analytic results can be obtained that display different behavior from their counterparts in (linear) solar neutrino flavor evolution.  For more general cases, numerical solutions require extended numerical algorithms to achieve stable solutions, and these solutions exhibit standard chaotic behavior.

\end{abstract}
\maketitle

\section{Introduction}
Neutrinos are trapped by multiple scattering from baryons at nuclear matter densities in a supernova.  However, outside the neutrinosphere, which characterizes the boundary of the baryon-rich region, neutrinos can propagate outward as coherent waves.  There is an extensive literature aimed at describing neutrino transport outside supernovae where the scattering medium can include the neutrinos themselves as well as electrons and baryons~\cite{Duan:2010bg,Pantaleone1992128,Notzold:1988ik,PhysRevD.48.1462,PhysRevD.53.5382,PhysRevLett.89.191101,PhysRevD.74.105010,PhysRevLett.103.05110,PhysRevD.65.053011,0954-3899-34-1-004,PhysRevD.85.065008,PhysRevLett.104.191102,0954-3899-38-3-035201,PhysRevLett.108.231102}.  [For a review, see Ref.~\cite{Duan:2010bg} and references therein.]  

Recently concerns have been raised regarding numerical stability and/or sensitivy of the numerical results to geometric symmetry~\cite{PhysRevD.88.045031,PhysRevLett.111.091101,Duan:2014gfa}.  The purpose of this paper is to gain some insight into the results of three-dimensional simulations by means of a one-dimensional model.  The model retains the non-linear nature of neutrino-neutrino scattering, as well as some of the density variation of particle species found in the full three-dimensional picture.  It also permits a study of numerical algorithms beyond those typically used in the three-dimensional case.  In our analysis, we find that achieving numerical stability requires a higher order algorithm than what is normally employed in the three-dimensional case.  Once numerical stability is achieved, the solutions for all but the most symmetric cases exhibit the the standard criteria for chaotic behavior.

The model also provides a second insight with respect to the asymptotic flavor evolution of neutrinos far from a supernova where the neutrino density has fallen to zero.  It address the question whether neutrino flavors evolve in the same manner as they do when traveling in a static medium electrons outside the sun~\cite{Wolfenstein:1977ue,Mikheev:1986gs}.  It will be seen that the non-linear nature of supernova neutrino evolution leads to different configurations at large distances than one might expect from the solar neutrino case, which is linear in form.  

\section{Background: Neutrino Medium Refractive Index}
A refractive index describes coherent propagation of a particle in a medium of scatterers, for which plane waves travel with an effective wavenumber:
\begin{equation}
  \label{eq:1}
  \kappa = kn,
\end{equation}
where the index $n$ depends upon the forward scattering amplitude of the scatterer:
\begin{equation}
  \label{eq:2}
  n = 1 + {2\pi \rho\over k^2} f(0),
\end{equation}
and $\rho$ is the number density of the medium particles.~\cite{vandeHulst:1949}

This leads to an effective wave equation: for any direction fixed by a wavenumber $\k$, the spatial variation apart from the overall phase $e^{ikx}$ is given by
\begin{equation}
  \label{eq:21}
  {\partial\over\partial x} \psi = i K \psi,
\end{equation}
where
\begin{equation}
  \label{eq:22}
  K = {2\pi\rho\over k} f(0).
\end{equation}
$K$ thus plays the role of a Hamiltonian.

For neutrinos, the index becomes a matrix in flavor space.  The matrix contributions are diagonal for electron and baryon scatterers.  In the absense of mixing, the neutrino medium also contributes a diagonal matrix index.  For simplicity, we consider only two neutrino flavors $e$ and $x$ (e.g.\ assuming flavors $\nu$ and $\tau$ to be fully mixed).  In the presence of neutrino mixing, the matrix index contribution from each neutrino scatterer is diagonal a basis consisting of its mixed state plus its orthogonal partner.  The index then becomes a density matrix~\cite{Pantaleone1992128}.

The mean-free path,
\begin{equation}
  \label{eq:6}
  L \sim {1\over \rho \sigma_{\rm tot}}.
\end{equation}
provides a figure of merit by which to judge a calculation using the index alone: additional {\it non-coherent} contributions arising from inelastic scattering or elastic scattering far from forward-going directions start to be come significant only over distances comparable to $L$.

From Ref.~\cite{Duan:2006an}, we note that the maximum neutrino density outside the neutrinosphere is roughly $10^{32}$~cm${}^{-3}$.  The total elastic $\nu-\nu$ cross section (without exchange) is
\begin{equation}
  \label{eq:6a}
  \sigma = {G^2 s \over 4\pi}.
\end{equation}
For neutrino energies of 10~MeV, this corresponds to a mean-free path of order 50,000~km.  This means that a refractive index picture in which neutrinos travel through a medium of other neutrinos is an excellent approximation for the length scales of interest, typically hundreds of kilometers or less.  This observation is important for what follows: however difficult it may be to implement the equations of motion, they capture the dominant physics.  Put another way, under these density conditions, corrections to the refractive limit are very unlikely to alter the conclusions.

In the presence of an additional medium of electrons, the complete expression is
\begin{equation}
  \label{eq:47}
  K = K_\nu + K_e + K_m,
\end{equation}
where
\begin{equation}
  \label{eq:16}
  K_e = \sqrt{2} G \rho_e \left[\begin{array}{cc}
    1 & 0 \\
    0 & 0
   \end{array}\right],
\end{equation}
and
\begin{equation}
  \label{eq:15}
  K_m = {\Delta m^2\over 2 E} 
  \frac{1}{2}\left[\begin{array}{cc}
    -\cos2\theta & \sin2\theta \\
    \sin2\theta & \cos2\theta
    \end{array}\right].
\end{equation}
The coupled neutrino term receives a contribution from each scatterer with momentum $\p_2$:
\begin{equation}
  \label{eq:42}
  K_\nu = 2\sqrt{2} G \int d\p_2 \,  (1 - \p_1\cdot\p_2)
  \left[\begin{array}{cc}
   \rho_{\nu_e}(\p_2) & 0\\
    0 & \rho_{\nu_x}(\p_2)
    \end{array}\right],
\end{equation}
where $\p_1$ is the neutrino beam momentum.  In full calculations, the integral over $\p_2$ is weighted with a distribution function.

\section{Model Studies}
The focus of our study is a model with a single space-time variable $x\sim t$.
\begin{equation}
  \label{eq:23}
  {\partial\over\partial x} \psi(x) = i K \psi(x).
\end{equation}

We consider the evolution of two groups of neutrinos, labeled $A$ and $B$, respectively, and and associated wave functions $\psi$ an $\chi$, evolving under a single variable $x$:
\begin{eqnarray}
  \label{eq:24}
  {\partial\over\partial x} \psi &=& K^A \psi \\ \nonumber
  {\partial\over\partial x} \chi &=& K^B \chi,
\end{eqnarray}
where $K^A$ and $K^B$ are identical except for the nonlinear neutrino-neutrino interaction term:
\begin{eqnarray}
  \label{eq:25}
  K^A_{\nu} &=& G\rho_\nu \, \chi \otimes \chi^\dag; \\ \nonumber
  K^B_{\nu} &=& G\rho_\nu \, \psi \otimes \psi^\dag.
\end{eqnarray}
This model does not include the full effects of three-dimensional geometry, but retains the non-linear nature of the optics.

Full-scale simulations employ a thermal distribution of neutrinos.  In fact, a frequently discussed outcome of these simulations is the idea an energy swap.  For this model, we retain a minimal aspect of a distribution by following the evolution of two incoherent neutrinos in each group: $\psi_<, \psi_>, \chi_<, \chi_>$, corresponding to two distinct neutrino energies $E_<$ and $E_>$.

At $x=0$, each of the four neutrinos begins propagation as a flavor eigenstate, e.g.\
\begin{equation}
  \label{eq:26}
  \psi_e(x=0) = 
  \left[ \begin{array}{c}
      1 \\ 0
    \end{array}\right]
\end{equation}
for pure $\nu_e$, and
\begin{equation}
  \label{eq:27}
  \psi_x(x=0) = 
  \left[ \begin{array}{c}
      0 \\ 1
    \end{array}\right]
\end{equation}
for pure $\nu_x$.
We take $K_m$ as given in Eq.~\ref{eq:15}.  For $K_e$, we write
\begin{equation}
  \label{eq:28}
   K_e = \sqrt{2}G \rho_e F_e(x) {1\over2} 
  \left[ \begin{array}{cc}
      1 & 0 \\
      0 & -1
    \end{array}\right].
\end{equation}
In this study we have removed all matrix traces as they do not contribute to flavor oscillations.

 The form factor $F_e(x)$ accounts for the density falloff of electron density in three-dimensional space:
 \begin{equation}
   \label{eq:29}
   F_e(x) = ({R \over R + x})^2
 \end{equation}
where $R $ sets the scale of the neutrinosphere radius.
\begin{equation}
  \label{eq:30}
   K_\nu = \sqrt{2}G \rho_e F_\nu(x) \, \chi  \otimes \chi^\dag,
\end{equation}
where there is an implied incoherent sum over the two energies $E_<$ and $E_>$.  The form factor $F_\nu(x)$ accounts both for the spatial falloff of neutrino density (as for electrons) and the fact that a typical angle of incidence of neutrino pairs in the full three-dimensional problem approaches zero (see Eq.~\ref{eq:12}):
\begin{equation}
  \label{eq:12}
  (1-\p_1\cdot\p_2) \sim O\left({1\over r^2}\right).
\end{equation}
We then have
\begin{equation}
  \label{eq:31}
   F_\nu(x) = ({R \over R + x})^4
\end{equation}

The model has corresponding features to that reported in Ref.~\cite{PhysRevD.88.045031}.  That model also has a single variable, which can be considered to be one spatial dimension.  It has no electron or neutrino density variation, unlike the framework described above.  It also has fixed rather than variable neutrino densities.  Nevertheless, some conclusions are quite similar.

\subsection{Analytic Results} 
We examine here the possibility of a flavor evolution effect due to neutrino-neutrino interactions that is analagous to the MSW effect due to the neutrino-electron interaction.  In the MSW effect, solar electron neutrinos propagate initially through a medium of electrons whose density is high enough that its optical effect dominates over vacuum oscillations.  As the electron density decreases adiabatically with distance, the neutrino state gradually evolves into a mass-2 eigenstate via the non-crossing rule.  For supernova neutrinos, one could imagine that, for a neutrino population restricted to electron neutrinos only, these neutrinos would propagate in a medium dominated by electron flavor; as the electron neutrino density falls off, these neutrinos would evolve to mass-2 eigenstates.   {\it In fact, this does not happen: two pure flavor eigenstates evolve as they pass through each other to a final state containing the same mass eigenstate mixtures as their initial states.}  

We can demonstrate this by induction.  In the mass basis, a pure $\nu_e$ initial state is
\begin{equation}
  \label{eq:9}
  \psi_{\nu_e}(x=0) =
  \left[ \begin{array}{c}
      \cos\theta \\
      \sin\theta
    \end{array}\right].
\end{equation}
According to the claim, this state at finite $x$ has the form
\begin{equation}
  \label{eq:10}
  \psi_{\nu_e}(x) =
  \left[ \begin{array}{c}
      e^{i\alpha}\cos\theta \\
      e^{i\beta}\sin\theta
    \end{array}\right] .
\end{equation}
We now demonstrate that this form is preserved under a step $x\to x+h$.
\begin{equation}
  \label{eq:13}
  d\psi = i h (K_\nu + K_m) \psi.
\end{equation}
In the mass basis,
\begin{equation}
  \label{eq:17}
  K_m \psi = 
  \lambda \left[ \begin{array}{cc}
      -1 & 0 \\
      0 & +1
    \end{array}\right] 
  \left[ \begin{array}{c}
      e^{i\alpha}\cos\theta \\
      e^{i\beta}\sin\theta
    \end{array}\right] 
  = 
  \lambda\left[ \begin{array}{c}
      -e^{i\alpha}\cos\theta \\
      +e^{i\beta}\sin\theta
    \end{array}\right] .
\end{equation}
The neutrino-neutrino matrix is proportional to $\chi \chi^\dag$:
\begin{equation}
  \label{eq:33}
  K_\nu =\mu \chi \chi^\dag.
\end{equation}
For this case, the symmetry requires that $\psi$ and $\chi$ evolve in the same way.  This means that the density matrix $K_\nu$ simply identifies $\psi$ with eigenvalue +1:
\begin{equation}
  \label{eq:34}
  K_\nu \psi = +\psi.
\end{equation}
Hence we get
\begin{equation}
  \label{eq:35}
  d\psi = 
  i h \left[ \begin{array}{c}
      (\mu - \lambda) e^{i\alpha}\cos\theta \\
      (\mu + \lambda)e^{i\beta}\sin\theta
    \end{array}\right] .
\end{equation}
That is, the mass admixtures do not change in magnitude; only the relative phases change.  Unlike the case of MSW, the pure electron state never evolves into a pure mass-2 eigenstate, {\it even though the eigenvalues of the propagation matrix never cross at any point in their evolution.}  The variation of $K$ is adiabatic with respect to stellar dimensions; however, the non-linear term in the propagation matrix varies {\it at exactly the same rate as the wave function on which it operates,} so in that sense adiabaticity is lost.
Since the neutrino mass states eventually decouple, the asymptotic survival probability for a state originally produced as $\nu_e$ is
\begin{equation}
  \label{eq:37}
  P_e \mathop{\longrightarrow}_{x\to\infty} (\cos^4\theta+\sin^4\theta)
  = 1 - 2\cos^2\theta\sin^2\theta.
\end{equation}

The conclusions of this analysis can be readily extended for any case where
$\chi_< = \psi_<;\, \chi_> = \psi_>$ at $x=0$.

\subsection{Numerical Studies} 
For all but the most symmetric cases Eqs.~\ref{eq:24} and~\ref{eq:25} must be integrated numerically.  We use the following parameters:
\begin{equation}
  \label{eq:32}
  \Delta m^2 = .003~{\rm eV}^2;\, \theta = 0.1;\,
  \rho_\nu = 10^{32}~{\rm cm}^{-3};\, \rho_e = 10^{30}~{\rm cm}^{-3};\,
  R = 10~{\rm km}.
\end{equation}

\subsubsection{Integration Algorithms}
The nonlinear dependence of $K_\nu$ upon neutrino wave functions means that one must take extra care when integrating the equation of motion numerically, so as to avoid serious loss of precision and/or loss of unitarity (normalization of $\psi$ and $\chi$).  The method used in Ref.~\cite{Duan:2006an} uses what we denote {\it unitarized Euler} (UE).  For each step increment $h$, the equation of motion has the form
\begin{equation}
  \label{eq:4}
  \Delta\psi = i K  h \psi,
\end{equation}
where $\psi$ represents one of the two-component flavor spinors.  If $K$ is traceless, one can rewrite Eq.~\ref{eq:4} as
\begin{equation}
  \label{eq:7}
  \Delta\psi = i \Theta \vsigma\cdot\n \psi,
\end{equation}
To the same order in $h$, we can also write ({\it cf.}\ Eq.~42 of Ref.~\cite{Duan:2006an})
\begin{equation}
  \label{eq:8}
  \psi + \Delta\psi = e^{i\Theta \vsigma\cdot\n} \psi = (\cos\Theta+i\sin\Theta\vsigma\cdot\n)\psi,
\end{equation}
where
\begin{equation}
  \label{eq:11}
  \Theta = h \sqrt{|K_{ee}|};\, \vsigma\cdot\n = {h\over\Theta} K.
\end{equation}
Equations~\ref{eq:8} and \ref{eq:11} preserve unitarity at every step.  The overall numerical uncertainty is linear in $h$.

The UE method is very sensitive to small, likely unphysical, perturbations near $x=0$.  Accordingly, the authors of Ref.~\cite{Duan:2006an} added an extra contribution to $K_e$ that is very high near the neutrinosphere but falls off exponentially.  We include such a term for all numerical calculations:
\begin{equation}
  \label{eq:49}
  K_e' =10^6\times K_e (x=0) \times e^{-50 x / R}.
\end{equation}

We found for this model that we still could not always maintain numerical stability, and therefore sought an alternative.  One method that achieves a higher level of precision while maintaining unitarity is fourth-order implicit Gauss-Runge-Kutta (GRK)~\cite{Dieci:1994,MR0159424}. 

\subsubsection{Convergence}
The GRK method has a global convergence rate of $O(h^4)$ for step size $h$.
To test convergence to $O(h^4)$, we write the partial integration of one component of the coupled, linear differential equation at a particular point $x$ as
\begin{equation}
  \label{eq:61}
  F_h(x) = F_e(x) + A h^4,
\end{equation}
where $F_e(x)$ is the exact result and $Ah^4$ is the leading term in the expression for the error.
If we calculate $F(x)$ using various step sizes $h_i$, we can form ratios involving different $h_i$:
\begin{equation}
  \label{eq:62}
  \frac{F_{h_i}(x) - F_{h_j}(x)}{F_{h_j}(x) - F_{h_k}(x)} 
  = \frac{h_i^4 - h_j^4}{h_j^4 - h_k^4}.
\end{equation}
We can then take a ratio of ratios:
\begin{equation}
  \label{eq:63}
   \left(\frac{F_{h_i}(x) - F_{h_j}(x)}{F_{h_j}(x) - F_{h_k}(x)}\right)
  / \left(\frac{h_i^4 - h_j^4}{h_j^4 - h_k^4}\right) = 1.
\end{equation}

For cases where convergence could be achieved, this ratio was tested
for a variety of numerical runs using a sequence of $h$ values, typically in the 1-cm range, at points where the result showed significant fluctuation.  In each case tested, the ratio in Eq.~\ref{eq:63} was within one or two percent of unity.  A convergence at another power of $h$ (e.g.\ $h^2$) would
 not yield such a ratio for this range of step sizes.   In order to achieve this power-law convergence, some runs required up to 25 iterations to compute the implicit GRK integration coefficients at each step, together with an iteration precision of one part in $10^{15}$.

The UE method should have a convergence rate of $O(h)$.  A similar ratio can be constructed that is linear in $h$ in order to test this property..  For runs and regions where the results appear to be stable, the ratio is indeed close to unity. 

\subsubsection{Test Runs}
We discuss here a representative set of test runs for initial conditions that have varying symmetry between groups $A$ and $B$.

\begin{figure}[H]
\centering
\mbox{\subfigure{\includegraphics[width=3.4in]{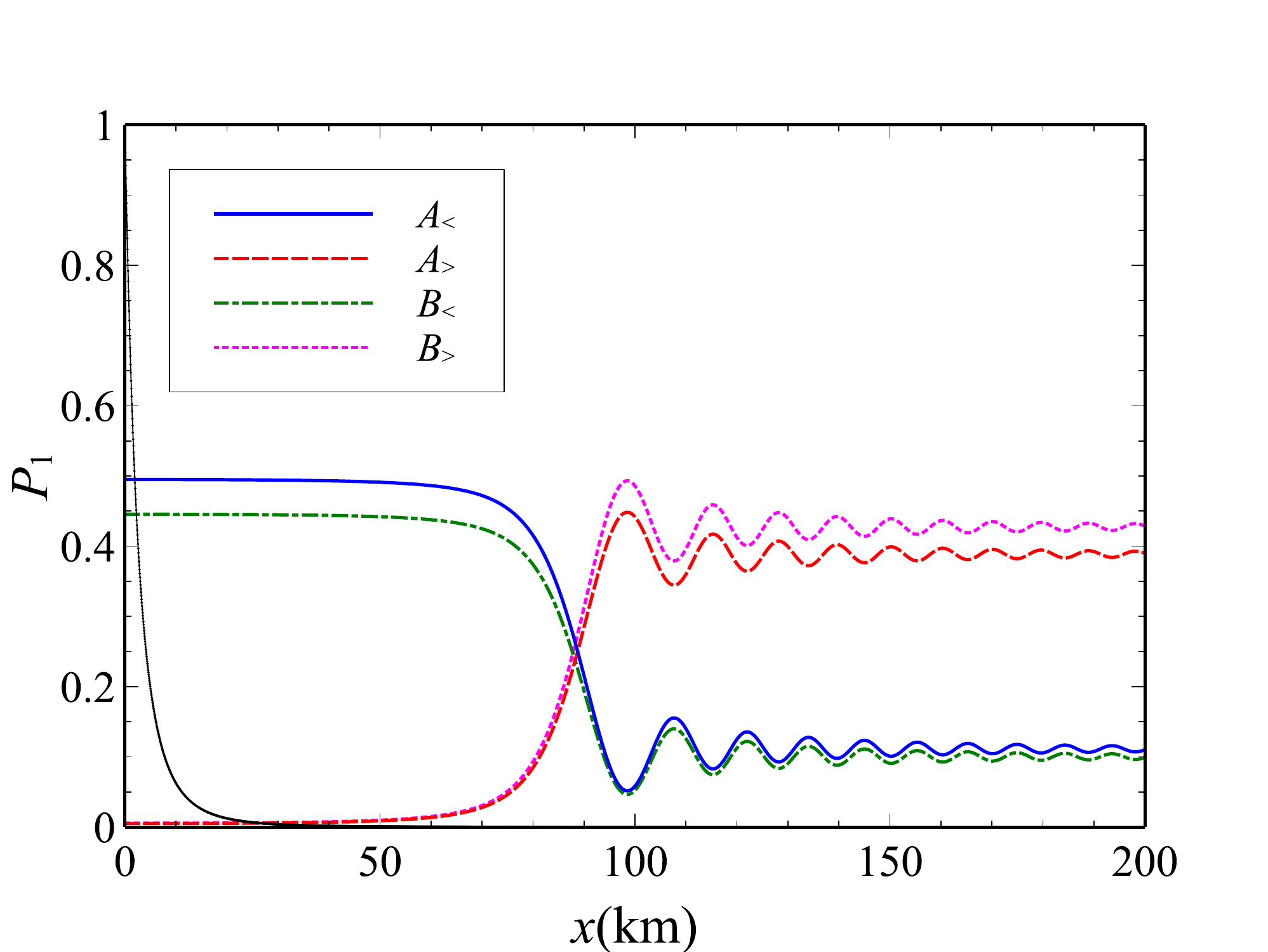}}\quad
\subfigure{\includegraphics[width=3.4in]{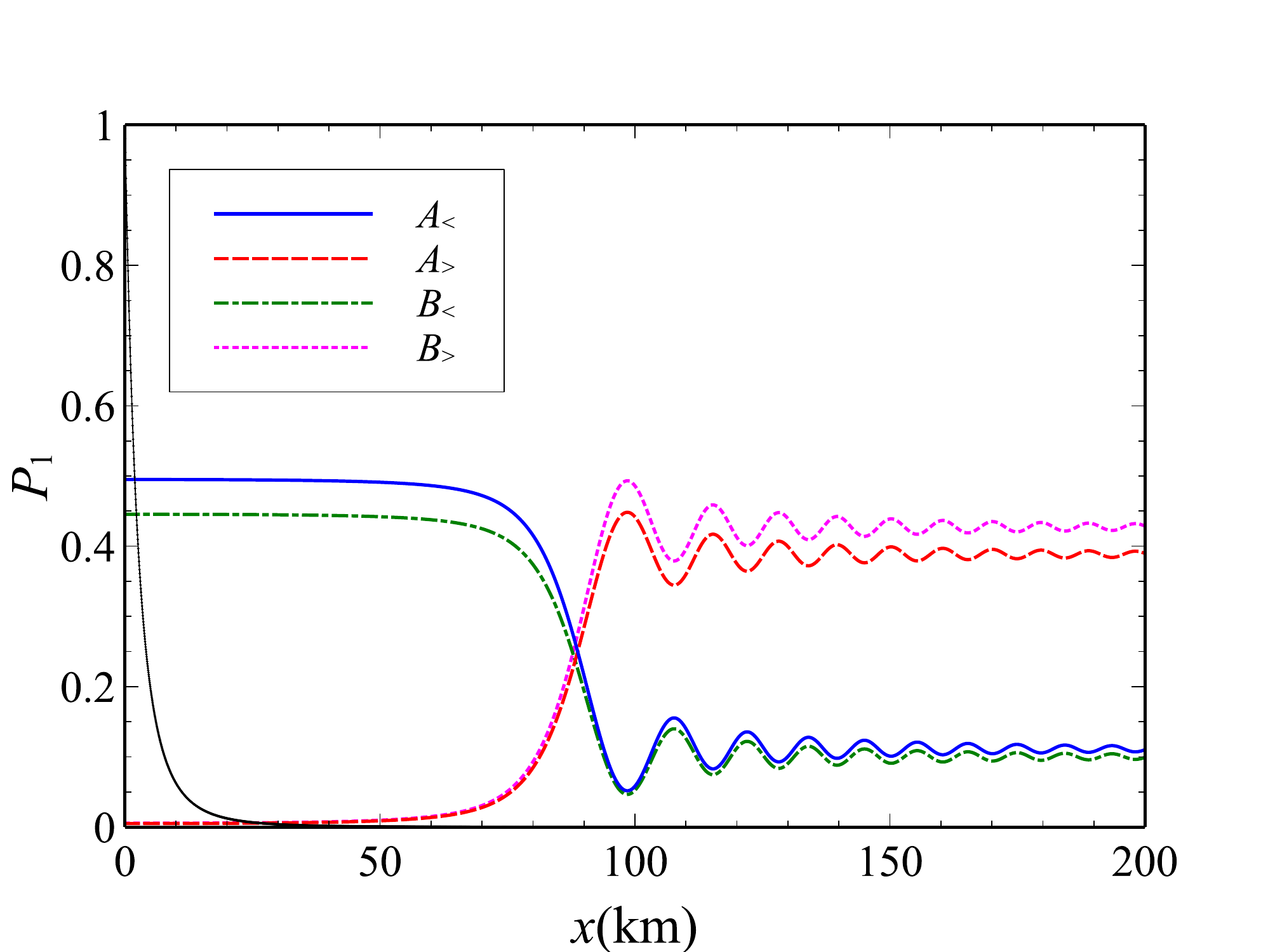} }}
\caption{Mass-1 emergence probability associated with each of four neutrinos as a function of $x$, with parameters of Case 1.  Left panel: GRK method; right panel: UE method} \label{fig:plot5050}
\end{figure}

\begin{figure}[H]
\centering
\mbox{\subfigure{\includegraphics[width=3.4in]{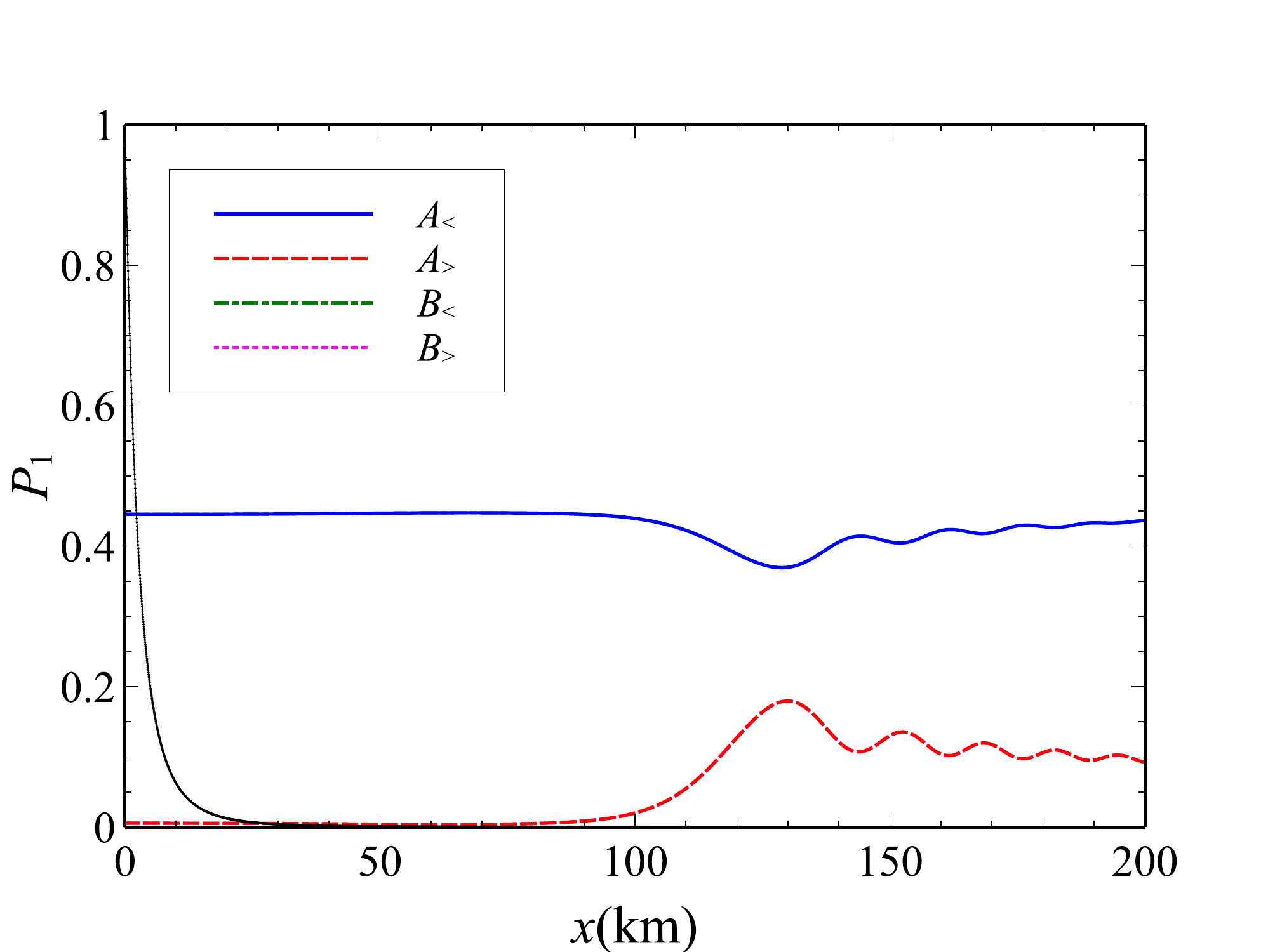}}\quad
\subfigure{\includegraphics[width=3.4in]{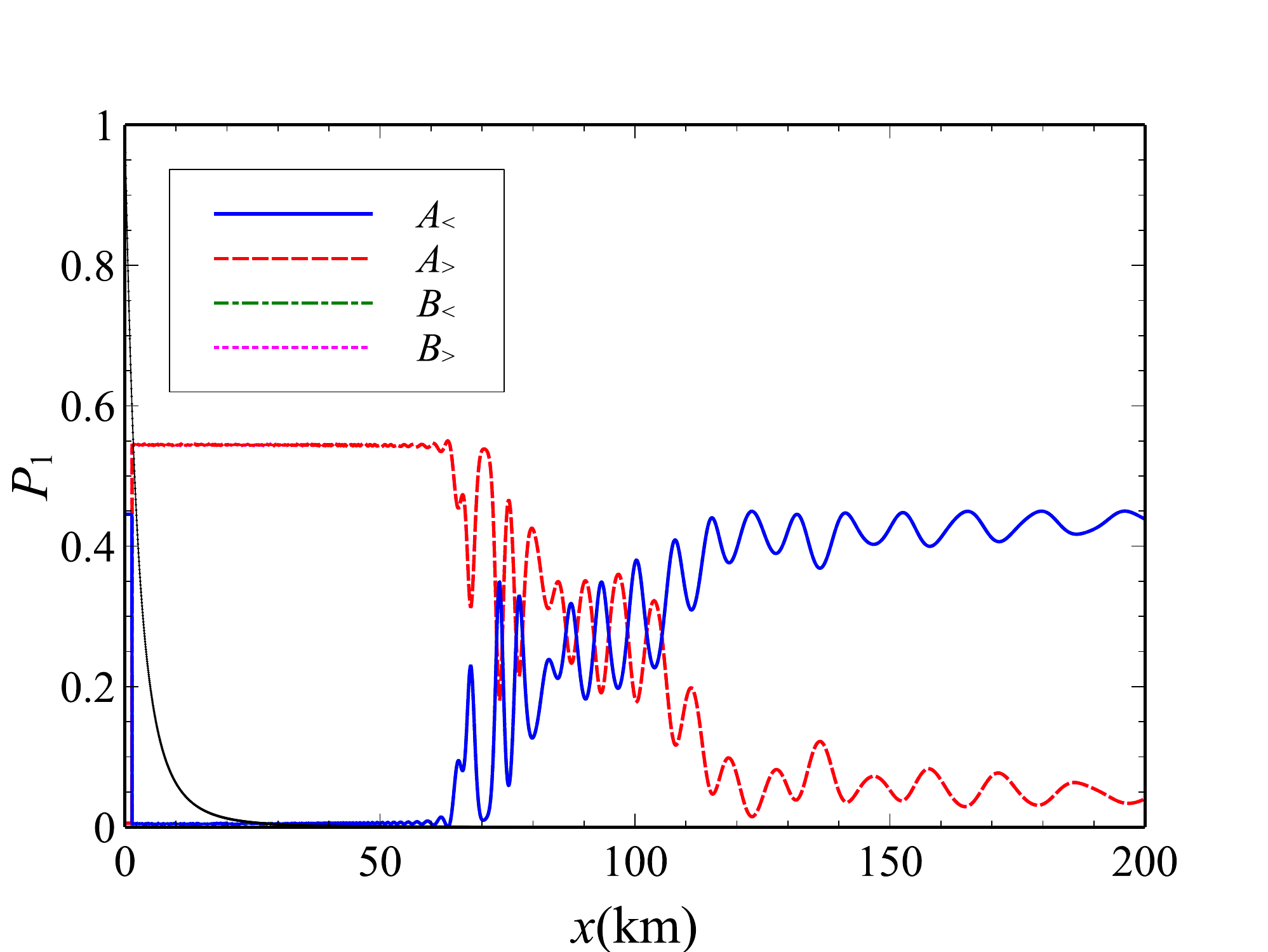} }}
\caption{Mass-1 emergence probability associated with each of four neutrinos as a function of $x$, with parameters of Case 2.  Left panel: GRK method; right panel: UE method.  In this particular case, $B_<=A_<;\, B_>=A_>$.} \label{fig:plot4555}
\end{figure}

\begin{figure}
\centering
\mbox{\subfigure{\includegraphics[width=3.4in]{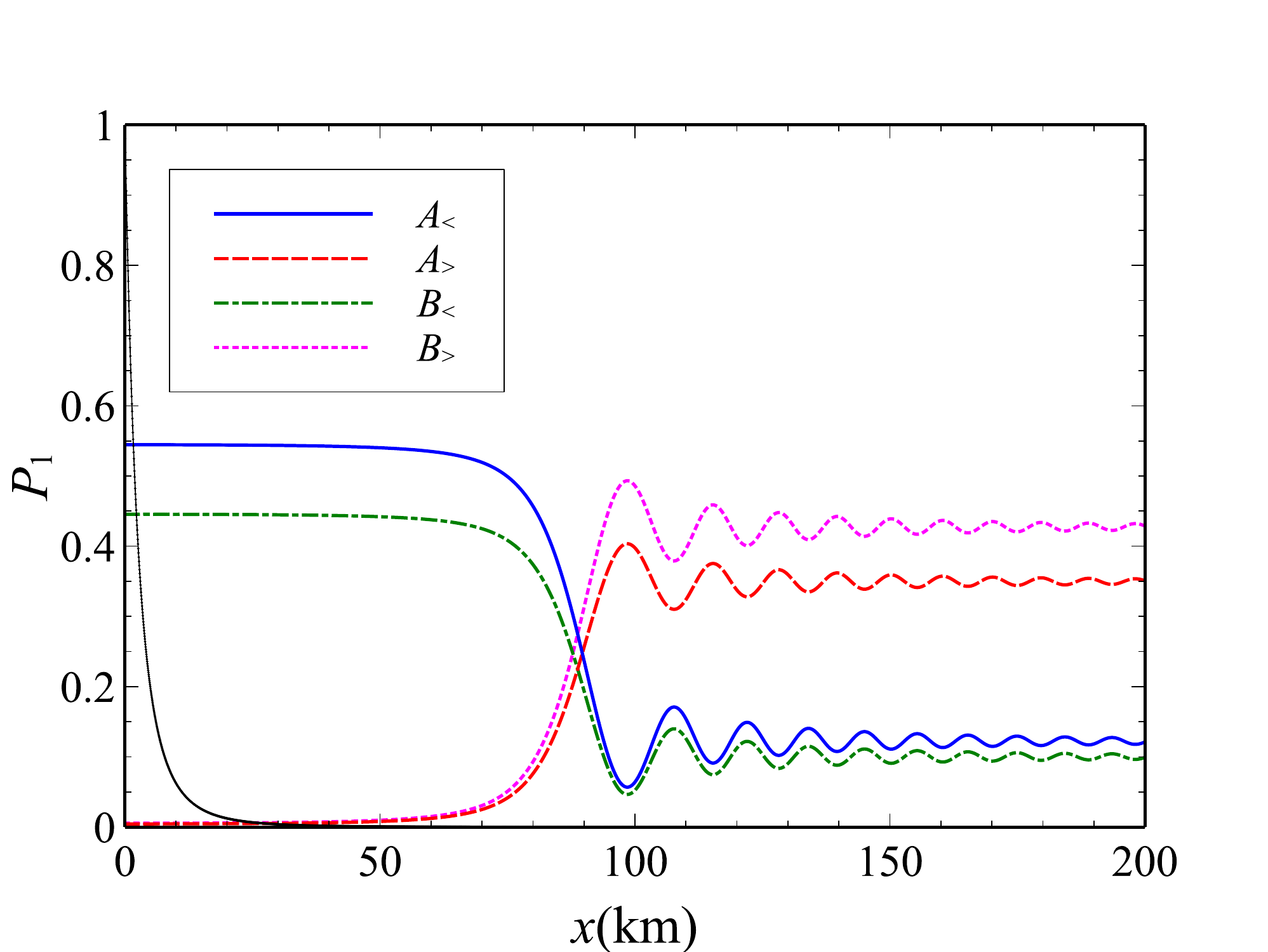}}\quad
\subfigure{\includegraphics[width=3.4in]{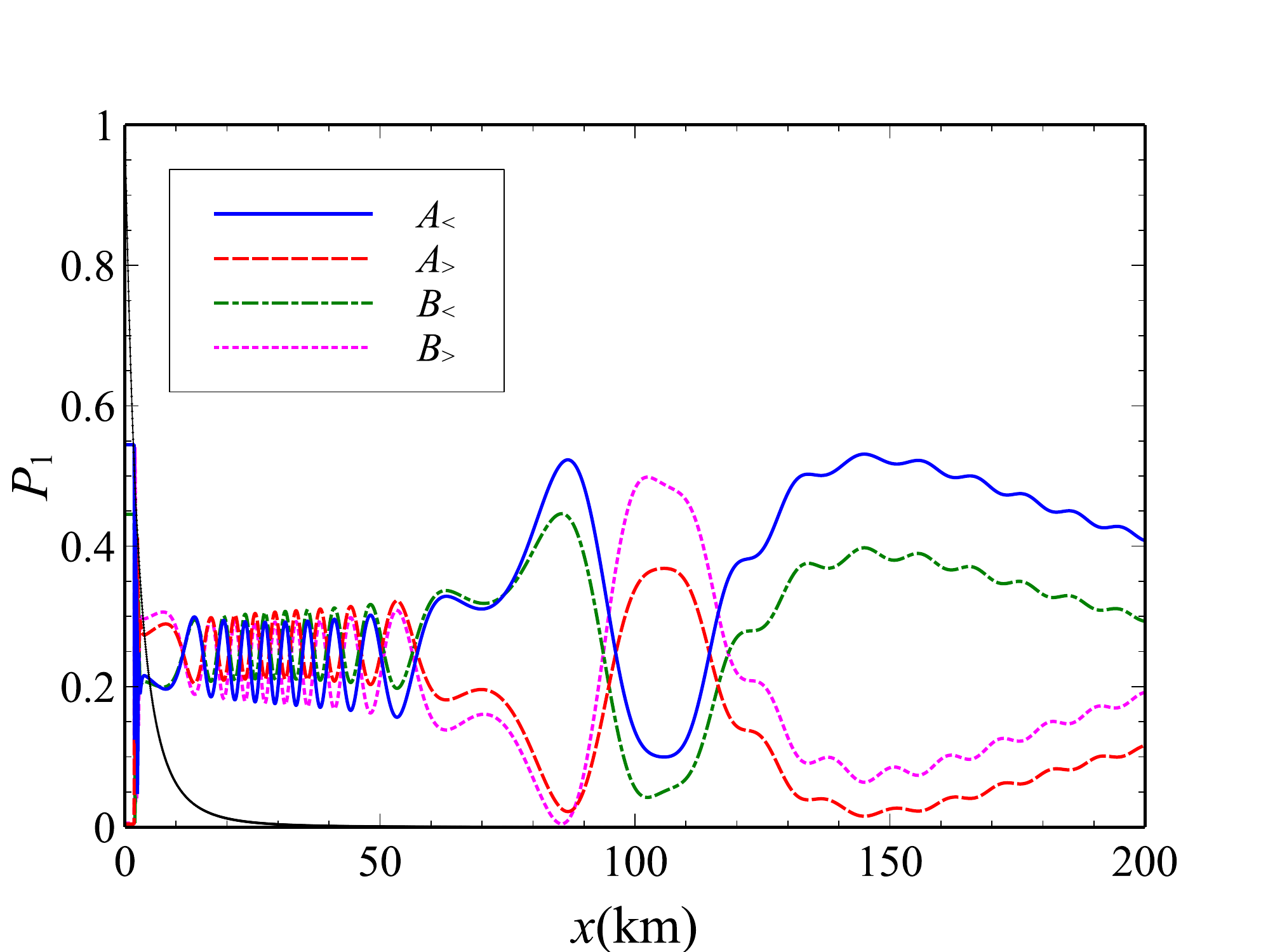} }}
\caption{Mass-1 emergence probability associated with each of four neutrinos as a function of $x$, with parameters of Case 3.  Left panel: GRK method; right panel: UE method} \label{fig:plot5545}
\end{figure}

\begin{figure}
\centering
\mbox{\subfigure{\includegraphics[width=3.4in]{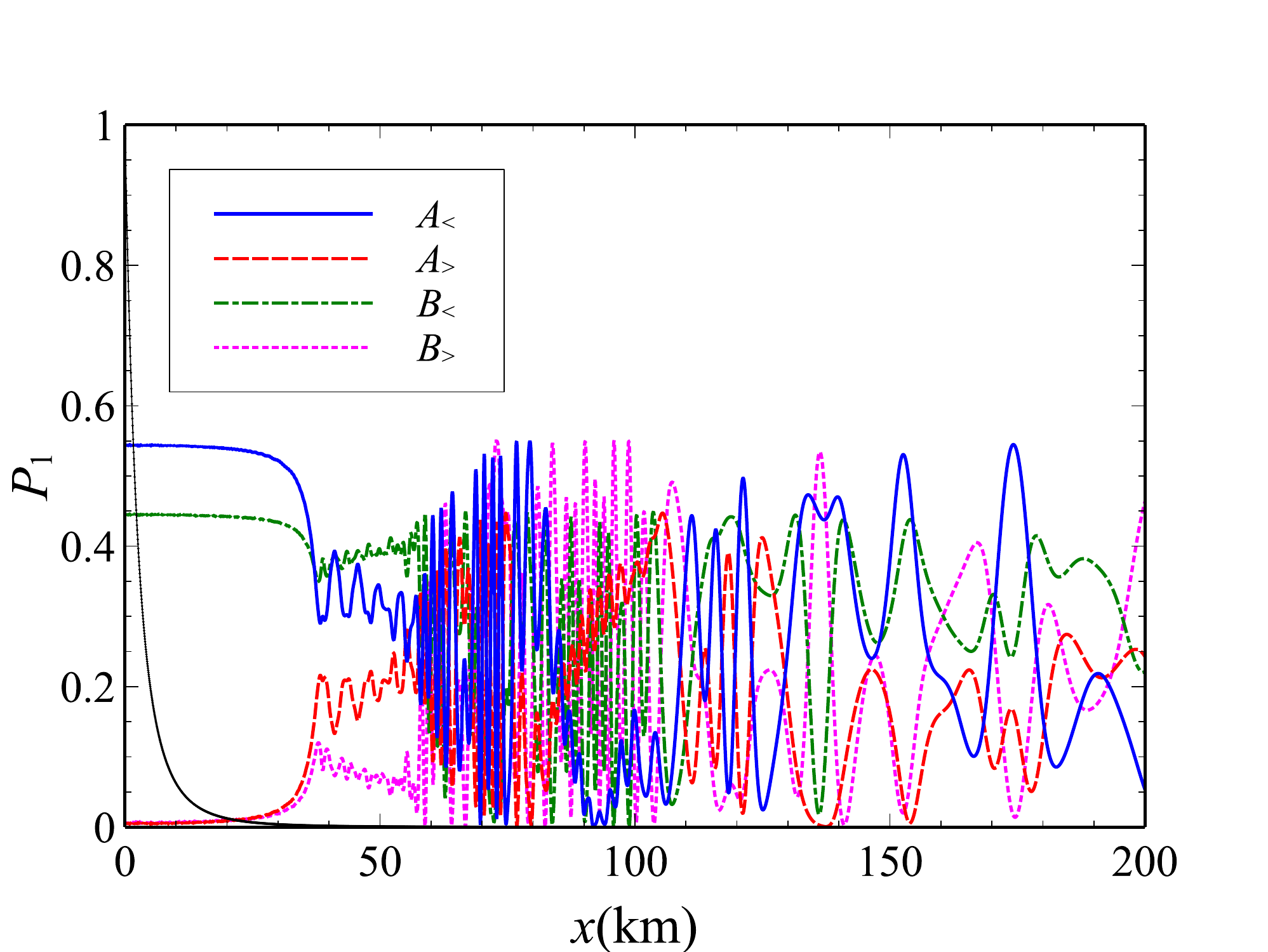}}\quad
\subfigure{\includegraphics[width=3.4in]{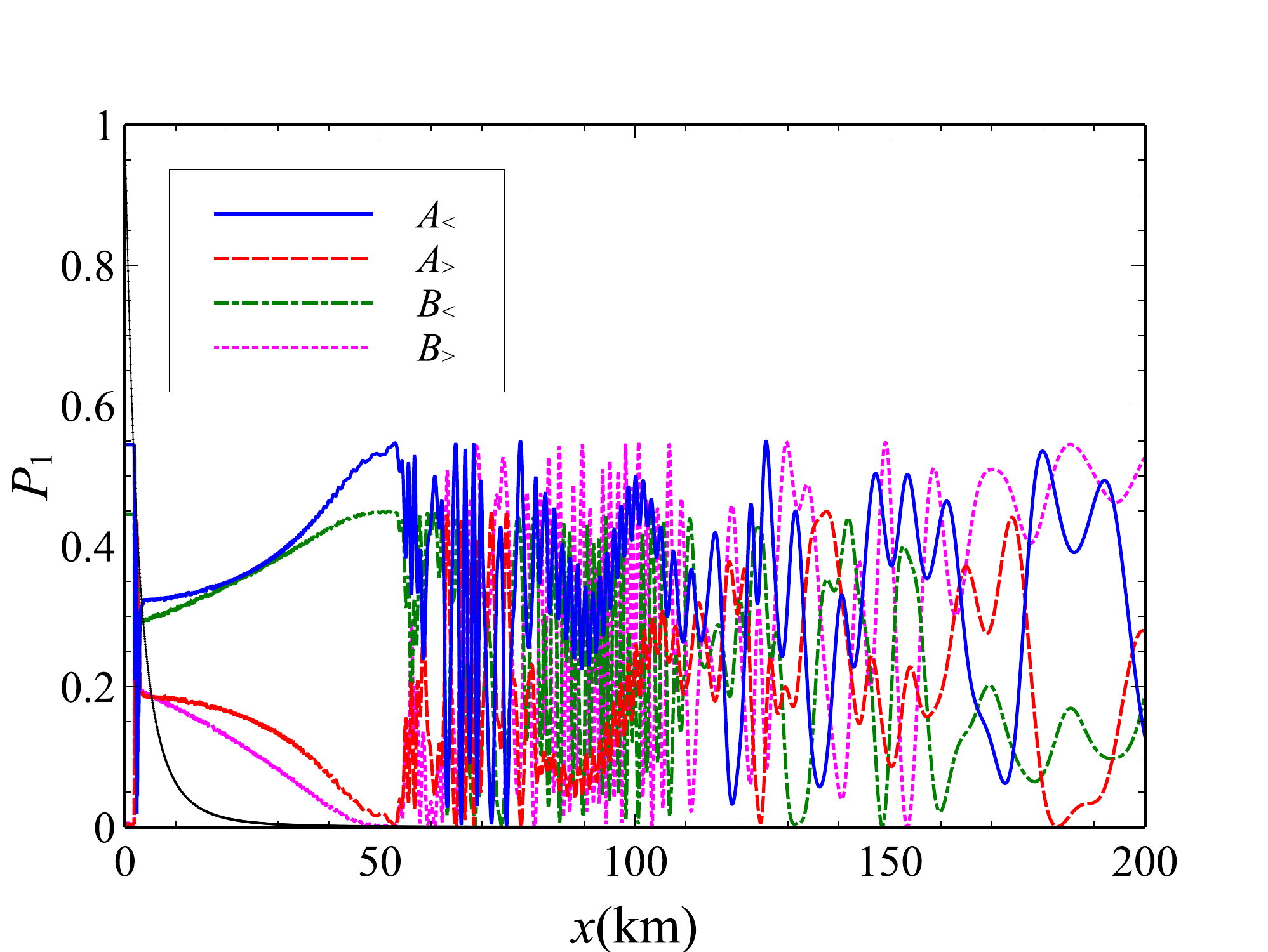} }}
\caption{Mass-1 emergence probability associated with each of four neutrinos as a function of $x$, with parameters of Case 4.  Left panel: GRK method; right panel: UE method} \label{fig:plot5545-x}
\end{figure}

\begin{itemize}
\item{\bf Case 1:}
For our initial case, we take both energies to be the same:
\begin{equation}
  \label{eq:84}
  E_< = E_> = 10~{\rm MeV}
\end{equation}
and
\begin{equation}
  \label{eq:91}
  |\psi_<^e|^2 = 0.5;\, |\psi_<^x|^2 = 0;\, 
  |\psi_>^e|^2 = 0;\, |\psi_<^x|^2 = 0.5;\, 
  |\chi_<^e|^2 = 0.45;\, |\chi_<^x|^2 = 0;\, 
  |\chi_>^e|^2 = 0;\, |\chi_<^x|^2 = 0.55.
\end{equation}
That is, equal populations of $\nu_e$ and $\nu_x$ in one direction pass through slightly differing populations of $\nu_e$ and $\nu_x$ in the other direction.  The results are shown in Fig.~\ref{fig:plot5050}.  In each case we plot the emergence probability of mass-1 neutrinos (choosing mass eigenenstates since those propagate over very large distances without further mixing).
In this case, both the GRK and UE integration methods converge, and yield the same results to $O(h^4)$ and $O(h)$, respectively.
\item{\bf Case 2:}
In this case, we use a two-energy spectrum:
\begin{equation}
  \label{eq:85}
  E_< = 9~{\rm MeV};\, E_> = 11~{\rm MeV}
\end{equation}
and equal neutrino populations $A$ and $B$:
\begin{equation}
  \label{eq:92}
  |\psi_<^e|^2 = 0.45;\, |\psi_<^x|^2 = 0;\, 
  |\psi_>^e|^2 = 0;\, |\psi_<^x|^2 = 0.55;\, 
  |\chi_<^e|^2 = 0.45;\, |\chi_<^x|^2 = 0;\, 
  |\chi_>^e|^2 = 0;\, |\chi_<^x|^2 = 0.55.
\end{equation}
The symmetry of the initial conditions means that both $A$ and $B$ neutrino populations will evolve in the same way.  The results are shown in Fig.~\ref{fig:plot4555}
In this case, the GRK method converges to $O(h^4)$.  The UE method does not converge at any value of $x$ after $x~1$~km.
\item{\bf Case 3:}
In this case, we gain use equal energies 
\begin{equation}
  \label{eq:86}
  E_< = E_> = 10~{\rm MeV}
\end{equation}
and a different population from that of Case 1.
\begin{equation}
  \label{eq:93}
  |\psi_<^e|^2 = 0.55;\, |\psi_<^x|^2 = 0;\, 
  |\psi_>^e|^2 = 0;\, |\psi_<^x|^2 = 0.45;\, 
  |\chi_<^e|^2 = 0.45;\, |\chi_<^x|^2 = 0;\, 
  |\chi_>^e|^2 = 0;\, |\chi_<^x|^2 = 0.55.
\end{equation}
In this case the flavor populations are unequal in both directions.
The results are shown in Fig.~\ref{fig:plot5545}
In this case, the GRK and UE integration method converges to $O(h^4)$ using step sizes of 0.5~cm or less, but the UE method does not converge even at these small step sizes.
\item{\bf Case 4:}
In this case, we use a two-energy spectrum
\begin{equation}
  \label{eq:87}
  E_< = 9~{\rm MeV};\, E_> = 11~{\rm MeV}
\end{equation}
and an asymmetric initial population:
\begin{equation}
  \label{eq:94}
  |\psi_<^e|^2 = 0.55;\, |\psi_<^x|^2 = 0;\, 
  |\psi_>^e|^2 = 0;\, |\psi_<^x|^2 = 0.45;\, 
  |\chi_<^e|^2 = 0.45;\, |\chi_<^x|^2 = 0;\, 
  |\chi_>^e|^2 = 0;\, |\chi_<^x|^2 = 0.55.
\end{equation}
This is the same as Case 2 except that the energies are unequal as well.
The results are shown in Fig.~\ref{fig:plot5545-x}
In this case, the GRK method converges to $O(h^4)$ until $x\sim 50~{\rm km}$, for which the rapid variations in the solution are confirmed to $O(h^4)$.  Beyond that range of $x$ it was difficult to obtain a stable solution.  In an attempt to improve convergence, we used a sixth-order implicit Runge-Kutta Algorithm~\cite{Dieci:1994,MR0159424}.  With this improved algorithm,  we were able to main stability with a precision of $O(h^6)$ up to about $x\sim75$ km, but could not move much further, even with step sizes $h\sim~1$ mm.  The UE method does not converge at any value of $x$.
\end{itemize}

\begin{figure}
\centering
\includegraphics[width=3.5in]{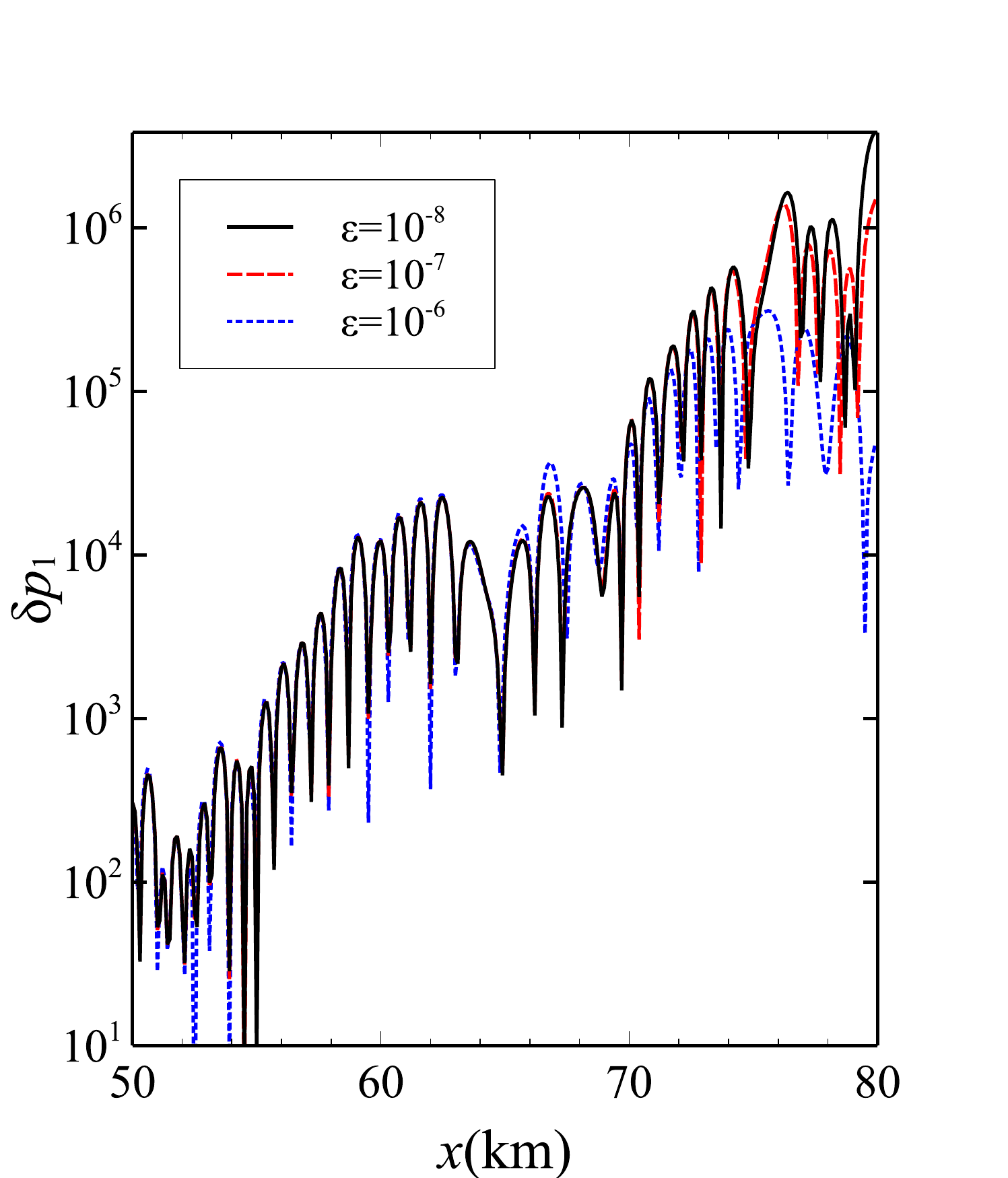}
\caption{Growth in relative separation of solutions for small initial displacements, where $\epsilon$ is defined in the text.  Curves are normalized to unity at $x=0$.} \label{fig:exponent}
\end{figure}
Further investigation of Case 4 suggests that the behavior is in fact chaotic.  To see this, we consider slightly modified parameter sets where
\begin{equation}
  \label{eq:52}
  |\psi_<^e|^2 = 0.55+\epsilon;\, |\psi_<^x|^2 = 0;\, 
  |\psi_>^e|^2 = 0;\, |\psi_<^x|^2 = 0.45;\, 
  |\chi_<^e|^2 = 0.45-\epsilon;\, |\chi_<^x|^2 = 0;\, 
  |\chi_>^e|^2 = 0;\, |\chi_<^x|^2 = 0.55.
\end{equation}
The displacement of initial condition $\delta|\psi|^2$ leads to a displacement of the outcome probability $\delta P_1$.  For a chaotic system, we expect to see~\cite{opac-b1078099}
\begin{equation}
  \label{eq:53}
  |\delta P_1(x)| \approx e^{\lambda x} |\delta P_1(x=0)|.
\end{equation}
In Fig.~\ref{fig:exponent} we show a calculated normalized displacement:
\begin{equation}
  \label{eq:54}
  \delta p_1(x) = \left| \delta P_1(x)\over \delta P_1(0)\right|
\end{equation}
for three choices of $\epsilon$: $10^{-8}$,  $10^{-7}$, $10^{-6}$.  By definition the curves begin at unity.  To the extent that Eq.~\ref{eq:53} is correct, the curves will lie on top of one another.  In practice, all three trajectories rise to $p_1 \sim 100$ and remain near that value until $x\sim 0$~km, after which they rise at an exponential rate.  For comparison, see Fig.~2.32 in Ref.~\cite{opac-b1078099}.

\subsection{Discussion}
From the examples described above, there are cases in the parameter space of initial conditions that lead to stable solutions using either GRK or UE methods, stable solutions using only the GRK method, and solutions that ultimate become chaotic, as confirmed by very high precision GRK algorithms.  That said, the cases yielding, convergent, non-chaotic solution represent very special instances in which some parameters are equal or satisfy a symmetry.  The more general case that one would encounter corresponds to Case 4, where the neutrinos have unequal energies, and the $A$ and $B$ $\nu_e$ and $\nu_x$ populations are also not the same.  The transition to a chaotic solution also occurs well before the neutrinos would be expected to reach their asymptotic mass contents.

These results are similar to those found in Refs.~\cite{PhysRevLett.108.231102,PhysRevD.86.125020,PhysRevD.88.073004,PhysRevLett.111.091101,PhysRevD.88.045031}, which examine sensitivities of solutions under very small variations of initial conditions.  In particular, the one-dimensional model of Ref.~\cite{PhysRevD.88.045031} produces very similar behavior to that found here.  That study found that results were stable for high neutrino densities.  Consistent with that study, we find that very high neutrino densities yield stable results.  However, we also find that as the neutrino density falls off, the results become chaotic.  Instabilities were also observed in a model study in Ref.~\cite{PhysRevD.72.045003}.  

In this study we distinguish between {\it numerical stability}, which can eventually be achieved to arbitrary precision using powerful integration algorithms, and chaotic behavior, which is a measure of the {\it physical sensitivity} of the results to initial condition.  As noted in these references, it is not clear whether a full three-dimensional approach using an algorithm that has proven convergence as a function of step size can yield stable results when initial conditions are asymmetric.  Even if this is possible, there remains the second issue of sensitivity, whose impact is physical rather than numerical.

\section{Summary}

With a one-dimensional model we have been able to explore several aspects of non-linear neutrino evolution outside a supernova neutrinosphere.  For certain limiting cases (such as a single-flavor initial state), one can obtain exact asymptotic results that differ from what one expects from the corresponding linear case (such as solar neutrino evolution).  For more general cases of initial flavor distributions, neutrino flavor (or mass) evolution becomes difficult to integrate numerically; one can achieve numerical convergence, but the solution and eventually exhibits the standard signals for chaotic behavior.  We note here that the framework of refractive optics remains valid even in the chaotic region.  The large neutrino mean-free path means that this system is in fact an excellent illustration of dynamical birefringence whose microscopic origin can be derived, whether or not a stable numerical solution exists.

\begin{acknowledgements}
The author wishes to thank A.~B.\ Balantekin, J.~F.\ Cherry, G.~M.\ Fuller, J.~H.\ Eberly, K.~E.\ Keister, and P.\ Marronetti for helpful discussions and correspondence.  This work was supported by the National Science Foundation through its employee Independent Research and Development Program.  The opinions and conclusions expressed herein are those of the author and do not represent the National Science Foundation
\end{acknowledgements}

\bibliography{SN_optics}

\end{document}